\documentclass[%
 aip,
 amsmath,amssymb,
 reprint,
 showkeys,
]{revtex4-2}

\usepackage{graphicx}% Include figure files
\usepackage{dcolumn}% Align table columns on decimal point
\usepackage{bm}% bold math
\usepackage[version=3]{mhchem}
\usepackage{amsmath}
\usepackage{amssymb}
\usepackage{color}

\usepackage{hyperref}% add hypertext capabilities
\usepackage[mathlines]{lineno}% Enable numbering of text and display math
\newcommand{\ds}{\displaystyle}

\makeatletter
\newcommand*{\rightharpoonupfill@}{%
  \arrowfill@\relbar\relbar\rightharpoonup
}

\newcommand*{\leftharpoondownfill@}{%
  \arrowfill@\leftharpoondown\relbar\relbar
}

\newcommand{\xrightleftharpoons}[2][]{%
  \ensuremath{%
    \mathrel{%
      \settoheight{\dimen@}{\raise 2pt\hbox{$\rightharpoonup$}}%
      \setlength{\dimen@}{-\dimen@}%
      \edef\CA@temp{\the\dimen@}%
      \settoheight\dimen@{$\rightleftharpoons$}%
      \addtolength{\dimen@}{\CA@temp}%
      \raisebox{\dimen@}{%
        \rlap{%
          \raisebox{2pt}{%
            $%
            \ext@arrow 0359\rightharpoonupfill@{\hphantom{#1}}{#2}%
            $%
          }%
        }%
        \hbox{%
          $%
          \ext@arrow 3095\leftharpoondownfill@{#1}{\hphantom{#2}}%
          $%
        }%
      }%
    }%
  }%
}
\makeatother
\begin{document}

% Article Length Limits
% PRL	Letter 3,750 words

% TITLE
% Titles should be concise and informative, clearly stating the main findings of the manuscript. 
% Avoid using new terminology, hyperboles assessing the quality of the work (“precise,” “important,” 
% or “accurate”), proper nouns and brand names (name of equipment, people, or places), and coined words. 

\title{The physics of boundary conditions in reaction--diffusion problems} % Force line breaks with \\
%\thanks{}%

\author{Francesco Piazza}
\affiliation{Dipartimento di Fisica \& Astronomia, Università di Firenze and INFN sezione di Firenze, 
Via G. Sansone 1, 50019 Sesto Fiorentino (FI), Italy}
\email{Francesco.Piazza@unifi.it}
%\altaffiliation[On leave of absence from]{...}%Lines break automatically or can be forced with \\

\date{\today}%  It is always \today, today,

%==============================================================================================
\begin{abstract} % 600 characters max.
\noindent The use of fully or partially absorbing boundary conditions for diffusion-based problems has 
become paradigmatic in physical chemistry and biochemistry to describe reactions occurring in 
solutions or in living media. However, as chemical states may indeed disappear, particles cannot, 
unless such degradation happens physically and should thus be accounted for explicitly. 
Here, we introduce a simple, yet general idea that allows one to derive the appropriate boundary 
conditions self-consistently from the chemical reaction scheme and the geometry of the physical 
reaction boundaries. As an illustration, we consider two paradigmatic examples,
where the known results are recovered by taking specific physical limits. More generally,
we demonstrate that our mathematical analysis delivers physical insight that 
cannot be accessed through standard treatments. 
\end{abstract}
 % Use showkeys class option if keyword
                              % display desired
\maketitle

%\tableofcontents

%=================================================================================================
%\section{\label{sec:level1}First-level heading:\protect\\ 
%Introduction}

%\noindent In the year 1785, the eclectic Dutch scientist Jan Ingenhousz 
%reported~\cite{Ingenhousz:1785tm} 
%the first known observation of what, after the more systematic work of 
%Robert Brown some 40 years later~\cite{Robert-Brown:1828ux,Brown:1828ab},
%would become engraved in physics as Brownian motion.
%It is no exaggeration to state that statistical physics all started with
%the mathematical laws of diffusive motion. Beyond the much celebrated 
%contribution given by Einstein's seminal 1905 paper~\cite{Einstein:1905uc}, 
%much momentum came from the pioneering work of the 
%Polish physicist Marian von Smoluchowski~\cite{Gudowska-Nowak:2017tt}. 
%
\noindent The work on diffusion-limited coagulation of colloidal 
particles of the Polish physicist Marian von Smoluchowski~\cite{Smoluchowski:1916fk}
can be rightly considered as one of the fundamental pillars of modern 
molecular physical chemistry, and, as a matter of fact, of the entire 
biochemistry of living organisms.
Smoluchowski's demonstration of how a non-equilibrium boundary value problem
can be used to describe chemical reactions
in solution has been a scientific breakthrough back then, 
and is to date a far-reaching  achievement. 
Nonetheless, this powerful idea rests on the somehow physically unsettling 
idea of using sinks to model chemical reactions occurring upon diffusive encounters. \\
\indent Let $A$ and $B$ be two particles of
radius $R_A$ and $R_B$ and diffusion coefficients $D_A$ and $D_B$, 
respectively, and let the goal be to compute 
the typical rate of $A-B$ encounters. 
The reasoning goes as follows: If the concentration of one 
of the species, say $c_A$, is much smaller than the other, then 
the problem can be reduced to that of a single A surrounded by a
homogeneous fluid of $B$s. Moreover, if the latter are themselves
diluted enough to be considered as non-interacting, the whole problem 
reduces further to a two-body problem. Smoluchowski's brilliant 
intuition was that, if diffusion is the rate-limiting step, i.e. 
the reaction that occurs upon contact proceeds
much faster than the typical time of diffusive approach, one 
may consider a steady non-equilibrium diffusive flux of $B$ particles
disappearing instantly at the contact distance $R=R_A+R_B$. Then, the total flux 
is the sought rate.
Mathematically, one should solve the stationary diffusion 
equation for the concentration field of $B$s around one $A$, $c(r)$,
with {\em absorbing} boundary conditions at $r = R$
and constant bulk concentration at infinite $A-B$ separation, that is,
\begin{equation}
\label{e:BPSmol}
\nabla^2 c = 0, \qquad \text{BC:} \quad 
c(R) = 0, \lim_{r\to\infty} c(r) = c_\infty
\end{equation}
Integrating the current $J = -D \nabla c$ ($D=D_A+D_B$) 
over the reaction sphere gives
\begin{equation}
\label{e:Smolrate}
\left. 
k = 4 \pi D R ^2 \frac{d c}{dr}
\right|_R = 4 \pi D R c_\infty
\end{equation}
which is known as the Smoluchowski rate~\footnote{Strictly speaking, as shown 
by Szabo~\cite{Szabo:1989fk}, this results is correct only in the limit $D_A\to0$}. 
This result is simple 
and elegant and, indeed, it conveys correctly the essential physics of 
diffusion-limited encounters in diluted solutions. A closer inspection, however, 
reveals some unsettling troubles. \\
\indent An absorbing boundary condition is certainly a clever 
mathematical trick, but it implies nonetheless a somewhat mysterious 
mass annihilation. 
In fact, as pointed out by Collins and Kimball (CK) in 1949, the result~\eqref{e:Smolrate} 
cannot be correct if not every $B$ particle that reaches the 
reaction radius eventually reacts~\cite{Collins:1949aa}.
Based on this observation, CK surmised that a more physical boundary 
condition should be of the Robin (mixed) type, namely
\begin{equation}
\label{e:CKBC}
\left.
         4 \pi D R ^2 \frac{d c}{dr}
\right|_R =  k^\ast c(R)
\end{equation}
The rate constant $k^\ast$ (units of inverse concentration times inverse time) 
somehow gauges the finite probability of a reaction indeed occurring and interpolates
between reflecting ($k^\ast=0$) and absorbing   ($k^\ast\to\infty$) BC.
With the BC~\eqref{e:CKBC}, the stationary solution yields a total flux 
\begin{equation}
\label{e:CKrate}
k = 4 \pi D R ^2 \left.\frac{d c}{dr} \right|_R =  k_S c_\infty 
\left(
\frac{h}{1+h}
\right)
\end{equation}
where $k_S = 4 \pi D R$ and $h = k^\ast/k_S$. This result has been famously 
employed by Shoup and Szabo~\cite{Shoup:aa} to reinterpret the 
classical calculation  by Berg and Purcell~\cite{Berg:1977aa} 
for the rate  of irreversible binding of ligands to $N$ receptors 
on a cell's membrane. In this case, each receptor acts as an absorbing 
disk, so that $k^\ast \propto N$. The point, however, is that 
fully or partially absorbing 
boundary conditions seem  to be physically
plausible only if particles  do really disappear. For example, in a process of
intake followed by degradation of some nutrient, say by 
a spherical colony of algae~\cite{Sozza:2018aa}. 
By contrast, the use of fully or partially absorbing 
BCs to model binding appears more elusive. \\
\indent The slight physical discomfort caused by the use of
absorbing boundary conditions to model chemical transformations
resides in the fact that, by doing this, particles cannot be 
properly distinguished from chemical states. 
A clear illustration of this ambiguity are so-called annihilation
reactions, such as when a fluorescent molecule $B^\ast$ relaxes 
upon collision with a quencher $A$, i.e. 
$B^\ast + A \to B + A$~\cite{Szabo:1989fk}. In this 
case, the diffusing particles can exist in either the excited 
or relaxed state, with a transition in chemical space 
being promoted by a physical collision with the quencher. 
However, while the excited state disappears, the particle is never gone.
It merely changes color, so to say. \\ 
\indent The real problem seems that there is no  general rigorous method 
at present to derive self-consistently 
the appropriate boundary conditions  for mass diffusion coupled to chemical 
transformations among different states that may occur 
in the bulk and on some reactive manifold. Of course, the boundary conditions 
should ensure mass conservation if the chemistry at hand prescribes so.
%
%In particular, it would be desirable to start by 
%stating the full problem so as 
%the rate equations that describe all chemical reactions 
%occurring at the surface are 
%written down explicitly, while the diffusion 
%equation is merely supplied with physically transparent boundary
%conditions. Typically, these  only account for impermeability (if mass is 
%conserved overall) or  mass loss, if there are explicit chemical 
%reactions that prescribe so. 
In this paper,  we introduce a simple, yet powerful idea
that unveils a  self-consistent 
method to solve this problem in general. 
In the following, we illustrate this method in the case of  
two well-known problems. \\
%
%Moreover, we show how 
%many know results can be recovered as a result of extremely transparent 
%physical limits, which 
%
\indent It is instructive to start from the very text-book case of 
annihilation reactions, such as fluorescence quenching.
Let $c^\ast$ and $c$ denote the concentrations 
of excited and relaxed $B$ molecules, respectively. 
First of all, the minimal physically transparent scheme should 
be laid out in chemical space. This may take the following form:
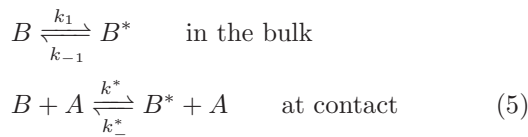
\begin{eqnarray}
\label{e:annihscheme}
&&B \xrightleftharpoons[k_{-1}]{k_{1}} B^\ast \qquad \text {in the bulk}\nonumber\\
&&B + A \xrightleftharpoons[k^\ast_{-}]{k^\ast} B^\ast + A \qquad \text {at contact}
\end{eqnarray}
The contact distance $r=a$ between the
particles and the quencher, $A$, is the reaction boundary. 
It can be seen that the reactions~\eqref{e:annihscheme}
describe a two-state system, where each molecule can switch between 
two states, either in the bulk or at contact with an $A$ molecule, but never disappears.  
Consequently, mass should be conserved,
which can be ensured through Dirac-delta reactivities~\cite{Szabo:1982vk}.
The reaction-diffusion equations corresponding to the chemical 
scheme~\eqref{e:annihscheme} read
\begin{subequations}
\label{e:annihphys}
% Eq. 1
\begin{eqnarray}
&&\begin{split}
\frac{\partial c}{\partial t} = D \nabla^2 c &+ k_{-1} c^\ast - k_1 c  \\
                                             &+ \frac{\delta (r-a)}{4 \pi r^2} 
                                             (k_-^\ast c^\ast  - k^\ast c ) 
\end{split} \nonumber \\
\label{e:annihphys:1}
&&\begin{split}
\frac{\partial c^\ast}{\partial t} = D \nabla^2 c^\ast &- k_{-1} c^\ast + k_1 c\\
                                             &+  \frac{\delta (r-a)}{4 \pi r^2} 
                                             {(k^\ast c - k_-^\ast c^\ast ) }
\end{split}                                               
\end{eqnarray}
% Eq. 2
\begin{eqnarray}
&&\lim_{r\to\infty} c = c_\infty \equiv 
                 \frac{k_{-1}}{k_1 + k_{-1}} c_0 \nonumber \\
&&\lim_{r\to\infty} c^\ast = c^\ast_\infty \equiv 
                 \frac{k_1}{k_1 + k_{-1}} c_0 \label{e:annihphys:2}
\end{eqnarray}
\end{subequations}
%
% Effective medium theory of rate processes among stationary reactive sinks
%~\cite{Cukier:1983tw}
%~\cite{Cukier:1985wh}
%
The whole idea is that the boundary conditions for the 
system~\eqref{e:annihphys:1} should follow self-consistently 
from the chemical reactions and take into account
that no mass ought to disappear in this problem.
This can be done by integrating eqs.~\eqref{e:annihphys} from $r=a-\delta a$ 
to $r=a+\delta a$ with $c(r) = c^\ast(r) = 0$ for $r < a$ (see supplemental
information).
This exercise, followed by the  limit $\delta a\to 0$, gives the desired BCs,
\begin{eqnarray}
4 \pi D a^2 \left. \frac{\partial c}{\partial r} \right|_a  -
                                   {k^\ast c(a) + k_-^\ast c^\ast(a)} = 0 
\label{e:BCannih1} \\
4 \pi D a^2 \left. \frac{\partial c^\ast}{\partial r} \right|_a  + 
                                     {k^\ast c(a) - k_-^\ast c^\ast(a)} = 0
\label{e:BCannih2}                                    
\end{eqnarray}
Note that these correspond to reflecting BCs of the total density $c+c^\ast$,
which states correctly that real particles can exist in either the relaxed or
excited state, but should not vanish mysteriously.\\
\indent In order to determine the encounter rate, we can restrict 
to the stationary problem. The general solution of this takes the form 

\begin{equation}
\label{e:annihgensol}
\begin{pmatrix} c \\ c^\ast \end{pmatrix} = 
\begin{pmatrix} c_\infty \\ c^\ast_\infty \end{pmatrix} + 
\frac{P}{r} \begin{pmatrix} 1 \\ \frac{k_1}{k_{-1}} \end{pmatrix}  + 
\frac{Q}{r} \begin{pmatrix} 1 \\ -1 \end{pmatrix} e^{-qr}  
\end{equation}
where $q = \sqrt{(k_1+k_{-1})/D}$. The constants $P$ and $Q$ are determined 
by the BC~\eqref{e:BCannih1} and~\eqref{e:BCannih2}, 
which gives the stationary profiles
\begin{eqnarray}
\label{e:annihsol}
&&\frac{c(r)}{c_\infty} = 1 - \frac{a}{r} \left( 1-
                                                e^{\beta \epsilon}
                                           \right)
                          \frac{\ds k^\ast e^{-q(r-a)}}{\ds k^\ast + k_-^\ast + k_S(1 + qa)} 
                \label{e:annihsol1}   \\
&&\frac{c^\ast(r)}{c^\ast_\infty} = 1 - \frac{a}{r} \left( 1-e^{-\beta \epsilon}
                                           \right)
                          \frac{\ds k_-^\ast e^{-q(r-a)}}{\ds k^\ast + k_-^\ast + k_S(1 + qa)} 
                \label{e:annihsol2}                 
\end{eqnarray}
The energy  $\epsilon = \beta^{-1} \log(k^\ast_- k_1/k^\ast k_{-1})$
quantifies the positive energy cost associated with increased non-equilibrium 
depletion of the excited species $B^\ast$
at the reaction surface when 
the boundary (i.e. contact with $A$ particles) acts as a quenching catalyst. 
Indeed, it can be seen that the profiles are both  flat at equilibrium, i.e.
when $k^\ast_-/k^\ast=k_{-1}/k_1$ or equivalently $\epsilon=0$. In this case, they 
equal the two bulk molar fractions everywhere, as expected.
Note also that we get the apparently 
surprising result $c(r)+c^\ast(r)=c_0$,
i.e. point-wise mass conservation. Indeed, this is the physically correct,
albeit rather uninformative, solution of the boundary 
problem for the total density with mass conservation and vanishing 
total flux at the reactive boundary.\\
\indent The exponential relaxation of the steady profiles conveys an important 
piece of physical information. As stated before, the whole theory rests on 
the hypothesis that $A$ molecules be diluted enough. By contrast to the standard
treatment, we now see clearly what {\em enough} means: the average $A$-$A$ separation should 
be greater than the typical relaxation length of the profiles, $q^{-1}$, 
i.e. $q c_A^{-1/3}\gg 1$.
This is tantamount to stating that the average time needed for a $B$ molecule to diffuse over 
the average $A$-$A$ separation, $\tau_A = D^{-1}c_A^{-2/3}$, should be longer 
than the average lifetime of the chemical states, $(k_1+k_{-1})^{-1}$.
In other words, demanding that particles be diluted in a reaction-diffusion system ought 
to involve a combination of diffusive relaxation time scales and 
lifetimes of chemical states.\\
\indent As we are explicitly distinguishing particles from chemical states, 
the correct way to compute the overall
reaction rate at the quenching surface is the following,
\begin{eqnarray}
\label{e:quenchrate}
k &=& J^\ast_- - J^\ast \equiv k^\ast_-c^\ast(a) - k^\ast c(a)  \nonumber \\ 
  &=& k_S (1 + qa) \frac{k^\ast_-c^\ast_\infty - k^\ast c_\infty}
                            {k^\ast_- + k^\ast + k_S(1 + qa)}
\end{eqnarray}
The stationary profiles~\eqref{e:annihsol1},~\eqref{e:annihsol2} 
and the rate~\eqref{e:quenchrate} 
convey a wealth of physical insight. First and foremost, it is now clear that the 
rate should be measured as an excess flow in chemical space 
relative to equilibrium. However, the  boundary 
condition~\eqref{e:BCannih1} shows that this 
is nothing but the total flux of excited molecules over the reactive boundary
according to the definition~\eqref{e:CKrate}.
Rather insightfully, our self-consistent theory reveals that the total particle 
flux equals the net current at the catalytic surface 
in chemical space. More generally, the sign of that current determines 
whether the catalyst acts as an active quencher or adds up to the excitation
in a non-equilibrium steady state, i.e. when 
$k^\ast_-/k^\ast \neq k_{-1}/k_1$. Indeed,
direct inspection of the profiles~\eqref{e:annihsol1} and~~\eqref{e:annihsol2}   
shows that one may have 
enrichment of either species at the reaction surface,
depending on the relative magnitudes of the relaxation-to-excitation rates ratio,
or equivalently on the sign of $\epsilon$.\\
\indent The familiar Smoluchowski setting is approached in the catalytic limit, 
i.e. when the relaxation rate at
the surface is much greater than  in the bulk. In the limit $k^\ast_- \to\infty$
one has 
\begin{eqnarray}
\label{e:quenchrateksm}
k \to k_S^q \equiv  k_S c^\ast_\infty (1 + qa) = 
k_S c^\ast_\infty 
\left( 
   1 + { \sqrt{\textrm{Da}} }
\right)  
\end{eqnarray}
where $\textrm{Da}=(qa)^2$ is the appropriate Damk\"ohler number for this problem.
We discover that this is greater than the expected Smoluchowski limit, $k_S c^\ast_\infty$.
In the conventional picture of particles pouring diffusively from infinity and 
vanishing into a sink, this may be interpreted as an extra accumulation of 
particles in the bulk, where reactions are not infinitely 
slow with respect to the typical diffusion time, $\tau_D = {a^2/D}$. Indeed,
the Smoluchowski rate is recovered in the infinite-diffusion limit,
i.e. $qa\to 0,\textrm{Da}\to 0$, or, equivalently, $k_1+k_{-1}\ll \tau_D^{-1}$.
%This may  be considered as the limit of infinitely slow equilibrium 
%in the bulk, that is, $k_1,k_{-1}\to0$ with $k_1/k_{-1}= const.$. 
The expression~\eqref{e:quenchrate} also yields the physically sensible result 
if the infinite-diffusion limit $qa\to 0$ is taken at finite 
surface quenching rate, $k^\ast_-$.  In this case one has
\begin{equation}
\label{e:quenchrateRL}
k = c_0 ( k^\ast_-k_1 - k^\ast k_{-1})
\end{equation}
which correctly only accounts for relative flow in chemical space.  \\
%
%Interestingly, the present  treatment of this chemical process allows one to access quantities 
%that are simply unavailable in the standard formulation. For 
%example, one may wish to quantify the efficiency $\eta$ of the surface quenching reaction 
%with respect to spontaneous relaxation occurring in the bulk. This can be computed as
%
%
%\begin{equation}
%\label{e:quenchrateRL}
%\eta \equiv \frac{c(R)/c^\ast(R)}{c_\infty/c^\ast_\infty} = 
%\frac{J^\ast}{J^\ast_-} e^{\beta \epsilon} 
%= 1 + \frac{ e^{\beta \epsilon}-1}{1 + \frac{k_1k_S(1+qR)}{k^\ast(k_1+k_{-1})}} 
%\end{equation}
%
%We see that the efficiency  of surface-catalyzed quenching is bounded 
%from above by $\eta_{\max} = e^{\beta \epsilon}$,
%which corresponds to the limit of either fast excitation at the surface, 
%$k^\ast/k_S \gg 1$, or fast relaxation in the bulk, $k_{-1}/k_1 \gg 1$,
%i.e. $c_\infty^\ast \ll c_0$.\\
%
\indent Let us now turn to another well-known problem, that of binding of a ligand
molecule to $\mathcal{N}$ receptors of size $b$ at the surface of a cell. 
The well-known Berg and Purcell (BP) 
formula~\cite{Berg:1977aa} has the form of eq.~\eqref{e:CKrate} 
and thus can be derived by solving a 
stationary diffusion problem subject to mixed BC, provided
$k^\ast = 4 D b \mathcal{N}$~\cite{Shoup:aa}.  
In other words, the surface reaction rate should correspond
to as many fully absorbing disks of radius $b$ as there are receptors on the cell's surface. 
Unfortunately, despite the correction worked out by Zwanzig 
through an elegant effective-medium argument~\cite{Zwanzig:1990aa}, the BP formula
still suffers from the same ambiguity connected to the use of sinks to describe binding reactions.
Following our method, this problem can be cast in a form that is more physically and biologically
transparent and that reduces to a BP-like formula in the appropriate physical limit. \\
\indent Let a cell of radius $a$ be covered with $\mathcal{N}$ receptors that 
can exist in their free form or in complex with some ligand. We consider the following 
scheme of reactions
%
%\begin{eqnarray}
%\label{e:LRscheme}
%&&L + R  \xrightleftharpoons[k_{-1}]{k_{1}} C \nonumber\\
%&&C \xrightarrow[]{k_d} R \qquad L \xrightarrow[]{k_d} \emptyset
%\end{eqnarray}
%
%
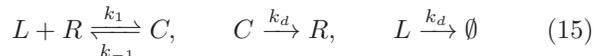
\begin{equation}
\label{e:LRscheme}
L + R  \xrightleftharpoons[k_{-1}]{k_{1}} C, \qquad 
C \xrightarrow[]{k_d} R, \qquad 
L \xrightarrow[]{k_d} \emptyset
\end{equation}
These correspond to complex formation/dissociation at the cell's surface 
and complex internalization 
followed by receptor recycling and ligand degradation, occurring with 
the same rate $k_d$.
Taking into account the diffusion of ligand molecules in the bulk, 
we can write the following set of equations,
\begin{subequations}
\label{e:cellprob}
% Eq. 1
\begin{equation}
\label{e:cellprob:1}
\frac{\partial \rho}{\partial t} = D \nabla^2 \rho  + 
\frac{\delta (r-a)}{4 \pi r^2} (k_{-1} N_C  - k_1 \rho N_R ) 
\end{equation}
% Eq. 2
\begin{equation}
\label{e:cellprob:2}
\frac{d N_C}{d t} = k_1 \rho(a,t) N_R - (k_{-1} + k_d) N_C
\end{equation}
% Eq. 3
\begin{equation}
\label{e:cellprob:3}
\lim_{r\to\infty} \rho(r) = \rho_0
\end{equation}
\end{subequations}
where $\rho(r,t)$ is the ligand concentration field, $\rho_0$ its bulk 
concentration and $N_R,N_C$ are the numbers of free receptors and complexes at the 
surface of a single cell, respectively. %Note that $N_L$ is a function of time. 
%The number of ligands at the cell's surface in  eq.~\eqref{e:cellprob:2} is
%$N_L = \rho(a,t)/\rho_c$, where $\rho_c$ is the volume concentration of cells. 
Combining eq.~\eqref{e:cellprob:2} with the conservation law
$N_R + N_C = \mathcal{N}$, one easily obtains the number of complexes in the 
stationary state, that is,
\begin{equation}
\label{e:BCstat}
N_C = \frac{\mathcal{N} \rho(a)}{\mathbb{K} + \rho(a)}
\end{equation}
%
%where $\mathbb{K} = \rho_c(k_{-1} + k_d)/k_1$ is the complex dissociation constant. 
where $\mathbb{K} = (k_{-1} + k_d)/k_1$ is the complex dissociation constant
(units of concentration). 
The appropriate boundary condition for the diffusion equation 
can be obtained again by integrating eq.~\eqref{e:cellprob:1} from $r=a-\delta a$ to $r=a+\delta a$
and then taking the limit $\delta a \to 0$. Taking into account eq.~\eqref{e:BCstat}, this 
yields the appropriate self-consistent BC, that is, 
\begin{equation}
\label{e:BCcell}
4 \pi D a^2 \left. \frac{d\rho}{dr}\right|_a - 
k_d \left( \frac{\mathcal{N} \rho(a)}{\mathbb{K} + \rho(a)} \right) = 0
\end{equation}
The stationary ligand profile outside the cell
has the form $\rho(r) = \rho_0(1 - \mu R/r)$, where 
$\mu$ is fixed by the BC~\eqref{e:BCcell},

\begin{equation}
\label{e:BCcellmu}
\mu = \frac{1}{2} \left( 
                              \frac{\mathbb{K}_d+\rho_0}{\rho_0} 
                               -\sqrt{
                                        \left( \frac{\mathbb{K}_d+\rho_0}{\rho_0} \right)^2 
                                        - \frac{4k_d \mathcal{N}}{k_S\rho_0}
                                      }
                   \right)
\end{equation}
with $\mathbb{K}_d = \mathbb{K} + k_d \mathcal{N}/k_S$.
%
%\begin{equation}
%\label{e:BCcellKd}
%\mathbb{K}_d = %\left( \frac{k_{-1}+ k_d}{k_1} \right) \rho_c 
%               \mathbb{K} + \frac{k_d \mathcal{N}}{k_S}
%\end{equation}
%
In this case, the setting is intrinsically a non-equilibrium one as in the standard 
Smoluchowski formulation, with the important difference that the present approach 
allows one to build in ligand binding and degradation in a physically and biologically 
transparent manner.
%The example discussed here should be considered as the 
%minimal model conceived to illustrate the generality of the present approach.\\
%
The overall ligand intake rate follows immediately as 
$k_i = 4\pi D a^2 (d\rho(a)/dr) = \mu k_S \rho_0$.
This result should be considered as the minimal extension of the Berg-Purcell 
formula in the direction of physical transparency of the underlying 
biochemical process. Once more, taking specific limits turns out to be insightful.\\
\indent At first glance, one may expect that the Smoluchowski rate $k_S\rho_0$ should 
be recovered in the limit of infinite degradation, $k_d\to\infty$. In fact, the 
maths reveal a slightly different physical picture, that is,
\begin{equation}
\label{e:BCcellratelim}
\lim_{k_d\to\infty} k_i = k_S \rho_0 \left( \frac{\kappa^\ast}{\kappa^\ast + k_S} \right)
\end{equation}
where $\kappa^\ast = \mathcal{N} k_1$. We thus learn that the BP formula 
corresponds to the limit of infinite internal degradation of ligand molecules
when the kinetics of complex formation at the membrane is properly accounted for. 
Moreover, the intuitive prescription $\kappa^{\ast}\propto \mathcal{N}$ posited 
by Shoup and Szabo~\cite{Shoup:aa} is  recovered here self-consistently.
Interestingly, our calculation unveils the correct  interpretation of the {\em intrinsic}
reaction rate $\kappa^\ast$, beyond the physical ambiguity of circular sinks.
Not surprisingly, in this minimal model it is the ligand-receptor 
association rate constant, $k_1$, that determines how 
effective a ligand absorber is a receptor-covered cell. 
Of course, if the  kinetics of 
ligand binding and the degradation chain were described in more detail, the calculation 
would provide more insight into how the inner workings of the whole biochemical
pathway should enter the overall intake rate. \\
%
%Perhaps a little more surprisingly, 
%we also discover that $\kappa^\ast \propto \rho_c^{-1}$. This simply means that $n$ cells 
%with similar numbers of receptors are $n$ times less effective individually  
%at ligand degradation~\footnote{This conclusion strictly holds provided $\rho_0 \gg \rho_c$,
%which allows one to describe ligand diffusion and binding in the presence of
%a single cell in the first place.}, which is the signature of mass-action law, encoded
%in the kinetics of ligand binding at the cells' membranes.\\
%in eq.~\eqref{e:cellprob:2}. 
%It sounds physically plausible 
%after all, but it should be noted that 
%this information is  unavailable in the BP formula. \\
%
\indent In summary, in this paper we have introduced a simple, yet powerful, idea that allows one to address
problems in molecular kinetics in solution on totally new grounds. Instead of relying 
on the old idea of supplementing the diffusion equation with empirical  
fully or partially absorbing  boundary conditions, we have shown that it is possible to derive 
the appropriate BCs self-consistently from the geometry of the reaction 
boundary and the details of the chemical reaction network. In our scheme, 
this can comprise reactions occurring in the bulk and at the reaction surface alike. 
The extent of the proposed change of perspective
can be readily appreciated from the potential applications of our idea. 
For example, with reference to ligand binding at a cell's membrane, 
the biochemical kinetic network considered
may be enriched by including different receptors, other surface proteins
that activate/inhibit them, more complex steps associated with internalization 
and recycling and a more detailed description of the degradation chain.
In principle, our method should provide a rigorous microscopic 
strategy to investigate how combined ligand diffusion and 
binding regulate any signaling pathways.
These include signaling cascades related to the emergence of polarisation
in cells steered by non-equilibrium self-generated gradients~\cite{Tweedy:2020vv}
created close to their plasma membrane through mechanisms such as the 
scheme~\eqref{e:LRscheme}. These gradients can in principle be measured experimentally and 
compared with our predictions based on detailed models of the underlying 
biochemical networks. \\
\indent More generally, our idea may be readily employed to investigate 
time-dependent  reaction diffusion problems too, as well as complex non-spherical 
reaction boundaries or even multiple disconnected boundaries arranged according 
to some pattern~\cite{Galanti:2016abc}. The latter perspective appears especially intriguing, 
as our method would allow one to explore how neighbouring reaction surfaces,
e.g. clusters of cells, influence each other with respect to a given microscopic kinetic (signaling) scheme
within a physically and biologically transparent model.

%Finally, it would be intriguing to test our idea in the presence of 
%nonlinearities, such as for catalytic/autocatalytic reactions.

\begin{acknowledgments}
\noindent The main idea this paper hinges upon occurred to the author as a result 
of a question asked by Alberto Parola on the physical meaning of absorbing 
boundary conditions in the Smoluchowski problem. 
%The author is indebted to
%him for demonstrating that questions are typically more insightful than answers. 
The author is also grateful to Paolo De Los Rios and Giuseppe Foffi for their 
precious comments.
\end{acknowledgments}

%\bibliography{bibl-diffusion,bio-11,DFT-FMT,tesi_biblio,bib_igg,
%              physics_of_boundaries,crowding,diff_obstacles,Refs,nanoreactors,scibib}


\end{document}